\documentclass[twocolumn,pre,showpacs]{revtex4}

\usepackage{epsfig}

\usepackage{amsmath}
\usepackage{amsfonts}
\usepackage{amssymb}

\begin{document}

\title{Physical realizability of small-world networks.}
\author{Thomas Petermann}
\altaffiliation{Present address: Unit of Neural Network Physiology,
Laboratory of Systems Neuroscience, National Institute of Mental
Health, Bethesda, Maryland 20892}
\email{Thomas.Petermann@alumni.ethz.ch}
\author{Paolo De Los Rios}
\affiliation{Institute of Theoretical Physics, LBS, Ecole
Polytechnique F\'ed\'erale de Lausanne - EPFL, CH-1015 Lausanne,
Switzerland}
\date{\today}

\begin{abstract}

Supplementing a lattice with long-range connections effectively
models small-world networks characterized by a high local and global
interconnectedness observed in systems ranging from society to the brain.
If the links have a wiring cost associated to their length $l$,
the corresponding distribution $q(l)$ plays a crucial role.
Uniform length distributions have received most attention despite
indications that $q(l) \sim l^{-\alpha}$ exist, e.g.\ for integrated
circuits, the Internet and cortical networks. While length distributions
of this type were previously examined in the context of navigability,
we here discuss for such systems the emergence and physical realizability
of small-world topology. Our simple argument allows to understand under which
condition and at what expense a small world results.

\end{abstract}
\maketitle

The explosion of research activity in the field of complex networks
has led to a novel framework in order to describe systems in
disciplines ranging from the social sciences to biology \cite{cnet}.
One feature shared by most real networks is the {\it small-world}
(SW) property involving a high degree of interconnectedness both at
a local and global level. That is, for every node, most nodes close
to it should also be close to each other and every pair of nodes is
separated, on average, by only a few links \cite{watts}. More
precisely, the latter is usually expressed with an at most
logarithmic increase of the mean distance as a function of the
system size. Although the SW phenomenon has first been introduced in
a social context \cite{milgram}, it is also relevant for
communication and technological systems such as the Internet
\cite{pastoryook} or electronic circuits \cite{cancho}. Small-world
properties are of great relevance for communication systems: SW
networks are particularly efficient for message passing protocols
that rely only on the local knowledge of the network available to
each node \cite{kleinberg01}. It has also been pointed out recently
that SW networks could describe the architecture of neuronal
networks: {\it in vitro} neuronal networks \cite{shefi}, brain
functional networks \cite{eguiluz} as well as the cerebral cortex
\cite{sporns} exhibit SW features. In fact, the topology plays a
crucial role in a neural network, since the high local
interconnectedness gives rise to coherent oscillations while short
global distances ensure a fast system response \cite{lago}.

To model SW networks in Euclidean space,
one starts with a regular lattice which is highly
interconnected locally and then rewires every link (connecting nodes
A and B) with probability $p$, that is the edge between the vertices
A and B is replaced by a long-range connection (or {\it shortcut})
between the nodes A and C, C being chosen at random \cite{watts}.
Clearly, the short global distances are due to the presence of
shortcuts, and as described in more detail below, it is the aim of
this paper to investigate the physical realizability of a SW
network. In the above model, $p$ allows to interpolate continuously
between a fully regular ($p=0$) and an entirely random ($p=1$)
topology, the precise nature of this transition being discussed
below. If the shortcuts are merely added (without losing local
connections), no significant changes in the emergence of the SW
topology result. We therefore deal with the model where re-wiring is
not accompanied by edge removal.

In the original formulation of the SW model, which received most of
the attention \cite{swrev}, the length distribution of the shortcuts
is uniform, since a node can choose any other node to establish a
shortcut, irrespective of their Euclidean distance. Yet, new
interesting properties emerge if this condition is relaxed, for
example if the distribution $q(l)$ of connection lengths $l$ decays
as a power law, $q(l)\sim l^{-\alpha}$. The navigability in such a
small world, for example, depends on the decay exponent $\alpha$
\cite{kleinberg}, and the nature of random walks and diffusion over
the network is also affected \cite{blumen,newcitation}. It was even
conjectured that the fundamental mechanism behind the SW phenomenon
is neither disorder nor randomness, but rather the presence of
multiple length scales \cite{kasturirangan} in agreement with $q(l)
\sim l^{-\alpha}$. Here we establish the properties of the wiring
mechanism which allows to realize SW  networks, the improved
navigability being a consequence of the SW property.

Real SW networks are unlikely to be successfully modeled
according to Watts and Strogatz' recipe given above: if shortcuts
have to be physically realized, the cost of a long-range connection
is likely to grow with its length. Since nodes connected by
shortcuts can be at any Euclidean distance from each other, it turns
out that the amount of resources that they have to invest in their
connections grows linearly with the linear system size, and it is,
{\it a priori}, unpredictable. This is far from optimal for systems
composed by entities with limited resources ({\it e.g.} providers or
neurons). Indeed, local (single node) and global {\it wiring cost}
considerations are likely to be key factors in the formation of real
SW networks \cite{chklovskii,laughlin,klyacho,buzsaki,sporns2,
yook}. Regarding connection-length distributions $q(l) \sim
l^{-\alpha}$, such measurements were reported for systems created
through self-organization, design and evolution, namely for the
Internet \cite{yook}, integrated circuits \cite{payman}, the human
cortex \cite{schuez} and for regions of the human brain correlated
at the functional level \cite{eguiluz}. Some modeling effort taking
into account the constraint of wiring minimization has been made for
systems where the connection lengths are \cite{karbowsky} or are not
distributed according to a power law \cite{barth,kaiser,gastner},
and such length distributions emerge quite naturally when wiring
costs along with shortest paths are minimized \cite{mathias}.

In this work we re-analyze the SW phenomenon from a wiring cost
perspective, for networks in $D$ dimensions, built using a power-law
decaying distribution of shortcut lengths. We find, both
analytically and numerically, that $\alpha<D+1$ is the condition for
the emergence of SW behavior. We also found that the local
interconnectedness increases with $\alpha$ and, given a fixed total
wiring cost, networks with larger values of $\alpha$ are smaller
worlds.

Given a $D$-dimensional lattice of linear size $L$, consisting of
$N=L^D$ sites, subject to periodic boundary conditions, it shall be
supplemented with $pN$ additional connections whose lengths are
distributed according to $q(l) \sim l^{-\alpha}$ as follows: for
every link to be added, we first choose its length according to the
(one-dimensional) distribution $q(l)$ and then put it on the lattice
by randomly choosing one endpoint and the other at the drawn
distance $l$, such that no pair of sites is connected by more than
one additional connection.

\begin{figure}[t]
\includegraphics[width=8cm]{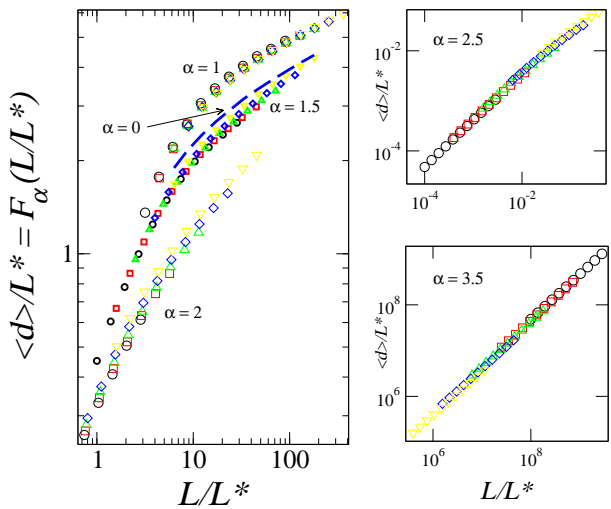}
\caption{Mean distance versus linear system size, both of these
quantities being rescaled by $L^*_\alpha(p)$, for $p=0.001$
($\circ$), $p=0.002$ ($\square$), $p=0.004$ ($\triangle$), $p=0.008$
($\diamond$) and $p=0.016$ ($\triangledown$). The exponent $\alpha$
ranges from 0 to 3.5 as indicated. The data collapses confirm Eqs.\
(\ref{equ:SW_cond}) and (\ref{equ:scal}) also for $\alpha>\alpha_c$
(lower right panel).} \label{fig:2D_coll}
\end{figure}

Clearly, a certain amount of real shortcuts, i.e.\ long additional
links, is required for SW topology to emerge \cite{watts,barth2}. It
can thus be anticipated that the exponent $\alpha$ has to be smaller
than a critical value $\alpha_c$. Before we give the argument
allowing to derive $\alpha_c$, let us recall that SW topology is
characterized by the following behavior of the mean distance
\begin{equation}    \label{equ:scal}
\langle d \rangle = L^* \mathcal F_\alpha \Bigl(\frac{L}{L^*}\Bigr),
\end{equation}
the scaling function obeying \cite{barth2,newman1}
\begin{equation}    \label{equ:scal_func}
\mathcal F_\alpha (x) \sim \begin{cases}
x \qquad\qquad &\text{if} \qquad x \ll 1,\\
\ln(x) \qquad &\text{if} \qquad x \gg 1. \end{cases}
\end{equation}
In other words, SW topology corresponds to a logarithmic increase of
the mean distance with the system size ($L \gg L^*$) whereas in a
{\it large world} (LW), i.e.\ if $L \ll L^*$, $\langle d \rangle
\sim L$. For $\alpha=0$, the critical length scale in
Eq.~(\ref{equ:scal}) is given by $L^*(p) \sim p^{-1/D}$
\cite{newman1,demenezes}. If $\alpha$ is positive, we shall derive
$L^*$ (as well as $\alpha_c$) through the following indirect
argument: We look at the probability that an arbitrarily chosen
additional link is a real shortcut, that is, that it spans the
lattice,
\begin{equation}    \label{equ:int_c}
P_c(L)=\int_{(1-c)L/2}^{L/2} q(l) dl,
\end{equation}
c being small but finite, and require (our ansatz) the expected
number of such connections to be of the order of 1 \cite{barth2}:
\begin{equation}    \label{equ:arg}
P_c(L) \bigl[p^*(L) L^D\bigr] \simeq 1.
\end{equation}
Here $p^*(L) L^D$ is the desired number of additional links,
implying the emergence of SW topology for $p \gg p^*(L)$. After
evaluating the scaling of (\ref{equ:int_c}), Eq.\ (\ref{equ:arg})
reads
\begin{equation}    \label{equ:SW_cond}
p^*(L) \sim \begin{cases}
L^{-D} \qquad\qquad &\text{if} \quad \alpha<1,\\
\ln(L)/L^D \qquad &\text{if} \quad \alpha=1,\\
L^{\alpha-D-1} \qquad\quad &\text{if} \quad \alpha>1.
\end{cases}
\end{equation}
Eq.\ (\ref{equ:SW_cond}) implies $L^*(p) \sim p^{-1/D}$ for
$\alpha<1$, i.e.\ the behavior of $L^*$ in this $\alpha$-range is
the same as that for $\alpha=0$. In the case $\alpha>1$, we have
$L^*(p) \sim p^{1/(\alpha-D-1)}$, thus becoming infinite at
\begin{equation}    \label{equ:alc-1}
\alpha_c=D+1.
\end{equation}
We therefore have two possible regimes for $\alpha<\alpha_c$ while
LW behavior prevails for $\alpha \ge \alpha_c$. Fig.\
\ref{fig:2D_coll} shows the rescaled mean distances as a function of
the rescaled linear system size for different values of $\alpha$ and
$p = 0.001, 0.002, ...,0.016$ in each set for the case $D=2$. The
observed data collapses for all the chosen values of $\alpha$
confirm Eq.\ (\ref{equ:SW_cond}) obtained by our simple argument as
well as Eq.\ (\ref{equ:scal}). We numerically verified Eq.\
(\ref{equ:scal_func}), especially in the limit $L/L^* \ll 1$, the
logarithmic tail of $\mathcal F_\alpha$ further being exhibited best
for small $\alpha$ \cite{footnote1}.

As outlined above, a SW network is also characterized by a high {\it
local} interconnectedness. This topological property can for example
be measured by the clustering coefficient $C$ which is the
probability that two nodes are connected, given that they share a
nearest neighbor. In contrast to Watts and Strogatz' model, our
initial lattices are characterized by $C=0$, but by increasing the
exponent of the link-length distribution, the degree of clustering becomes
orders of magnitudes larger than for random networks with the same number of
nodes and links.

\begin{table}[b]
\caption{Behavior of the moments of the shortcut-length distribution
as a function of the linear system size $L$ (for the ``adding''
procedure 1).}
\begin{ruledtabular}
\begin{tabular}{ccccc}
 & $0 \leq \alpha<1$ & $1<\alpha<2$ & $2<\alpha<3$ & $\alpha>3$  \\
\hline
$\langle l \rangle$ & $L$ & $L^{2-\alpha}$ & const & const \\
$\langle l^2 \rangle$ & $L^2$  & $L^{3-\alpha}$  & $L^{3-\alpha}$  & const
\end{tabular}
\end{ruledtabular}
\label{tab:moments}
\end{table}
Let us now examine the wiring costs, which were our prime motivation
to look at SW networks with power-law decaying link-length
distributions, and an important ingredient for real SW networks.
The moments $\langle l \rangle$ and $\langle l^2
\rangle$ play a crucial role as far as these costs are concerned.
Indeed, finite $\langle l \rangle$ and $\langle l^2 \rangle$
would allow for predictable costs for each node,
and consequently for a better design of the network constituents.
The total wiring cost $C_W =p L^D \langle l \rangle$ is also an
important quantity, its minimisation governing, for example, the
evolution of cortical networks \cite{laughlin}. We find for the
first two moments the scaling relations summarised in Tab.\
\ref{tab:moments}, the expressions for integer $\alpha$ being
modified by logarithmic corrections. In 2 dimensions, SW topology
can be realized even if $\langle l \rangle = \text{const}$ (that is,
for $2< \alpha < 3 = \alpha_c$) whereas this is not the case in 1
dimension where $\langle l \rangle$ becomes finite in the $L
\rightarrow \infty$ limit only above $\alpha_c=2$. Moreover, if
$D=3$, it is even possible to have $\langle l \rangle = \mathcal O
(1) = \langle l^2 \rangle$ while still being in the SW regime for $3
< \alpha< 4 = \alpha_c$. An appropriate choice of the parameters $D$
and $\alpha$ is thus the key to modeling networks which are both
efficient (SW topology) and economical (low wiring costs).

\begin{figure}[t]
\begin{minipage}[b]{.49\linewidth}
   \epsfysize= 4.6cm \epsfbox{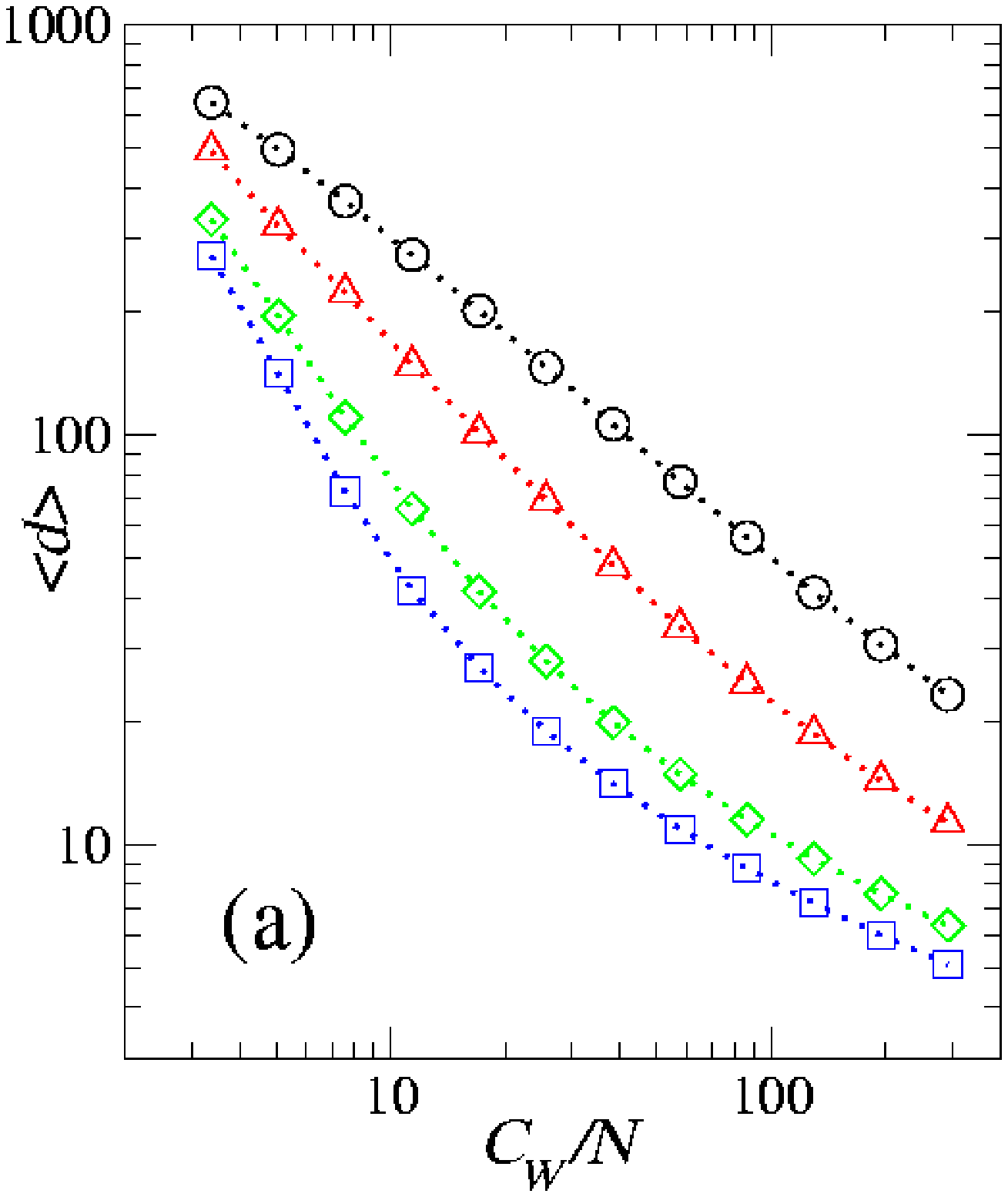}
\end{minipage}
\begin{minipage}[b]{.49\linewidth}
   \epsfysize= 4cm \epsfbox{md-2D.eps}
\end{minipage}
\caption{(a) Mean distance as a function of the total wiring costs
(divided by the number of sites) for 1 dimensional topologies
($N=10^4$). The curves ($\circ$: $\alpha=0$, $\triangle$:
$\alpha=1$, $\diamond$: $\alpha=1.5$ and $\square$: $\alpha=1.75$)
show that the mean distance decreases with $\alpha$ for a fixed
value of $C_W/N$. (b) Analogous results for $D=2$ [$N=500 \times
500$ and $\alpha=0$ ($\circ$), $\alpha=1$ ($\square$) and $\alpha=2$
($\diamond$)]. All the points shown here result from averaging over
100 realizations of networks.} \label{fig:wir_costs}
\end{figure}
It is furthermore interesting to have a closer look at the
relationship between the wiring costs and the topology. As $\alpha$
varies, one can ask what mean distance results given a total amount
of wiring length for the additional connections (i.e.\ the total
cost). Fig.\ \ref{fig:wir_costs}a reports these dependencies for
$\alpha=0,1,1.5$ and 1.75 (going from the uppermost to the lowest
set) for 1 dimensional topologies of $10^4$ sites. The largest value
of $\langle d \rangle$ (the leftmost circle) corresponds to the
length scale $L^*<10^3 \ll 10^4=L$, thus all the points in the
figure represent the system in the SW regime. It can clearly be seen
that the mean distance decreases with $\alpha$ at fixed wiring costs
$C_W/N$, i.e.\ the larger $\alpha$ the smaller the world. This
behavior is qualitatively recovered when expressing Eq.\
(\ref{equ:scal}) in terms of $x=C_W/N=p\langle l \rangle$. We made
similar observations in 2 dimensions (see
Fig.~\ref{fig:wir_costs}b).

Let us now point out the generality of our argument for the realizability of
SW networks in Euclidean space. In fact, it
also applies to a version of the SW model where the links are added
in a different way: at every site, a link is added with probability
$p$ - the other endpoint being chosen according to the
(D-dimensional) distribution $q(l)$ \cite{moukarzel}. This procedure
differs from the previous one in that the site from which the new
link will emanate ``sees'' the dimensionality of the lattice, giving
rise to a different normalization of $q(l)$ with respect to the
version treated above. Furthermore, the just described mechanism is
equivalent to adding a link between any pair of sites ${\bf x}$ and
${\bf y}$ with a probability proportional to $|{\bf x}-{\bf
y}|^{-\alpha}$ \cite{biskup}.

For this new construction procedure, the length distribution reads
$$
q(l)=\frac{l^{D-1-\alpha}}{\int_2^{L/2} \tilde l^{D-1-\alpha}
d\tilde l},
$$
where the factor $l^{D-1}$ explicitly accounts for the normalization
in D-dimensional space. With Eqs.~(\ref{equ:int_c}) and
(\ref{equ:arg}), which do not depend on the details of the
``adding'' mechanism, we obtain for the critical probability
$$
p^*(L) \sim \begin{cases}
L^{-D} \qquad\qquad &\text{if} \quad \alpha<D,\\
\ln(L)/L^D \qquad &\text{if} \quad \alpha=D,\\
L^{\alpha-2D} \qquad\quad &\text{if} \quad \alpha>D.
\end{cases}
$$
Conversely, this implies $L^* \sim p^{1/(\alpha-2D)}$ for
$\alpha>D$, and hence the existence of a SW regime as
long as
\begin{equation}    \label{equ:al_c-2}
\alpha_c < 2D
\end{equation}
in analogy with the previous reasoning. Inequality
(\ref{equ:al_c-2}) had already been derived \cite{moukarzel,biskup},
but in a less intuitive framework.

In summary, we have given a simple argument leading to the precise
conditions under which small-world topology emerges, and examined
the physical realizability of such networks. Due to the generality
of our argument, it is also applicable to other small-world models.
We further showed that small-world networks can be constructed in a
very economical way if the parameters $D$ and $\alpha$ are chosen
appropriately (although of course in real systems $D$ is seldom a tunable
parameter). As length distributions of the type investigated here
have been observed in a number of real-world networks, such as
integrated circuits, the Internet or the human cortex, we believe
this work to have intriguing implications in their modeling.

We thank Marc Barth\'elemy and Francesco Piazza for their valuable
comments, as well as to the EC-Fet Open project COSIN IST-2001-33555
and EU-FET Contract 001907 DELIS. Both the COSIN and DELIS contracts
have been supported through the OFES-Bern (CH).


\begin{thebibliography}{99}
\bibitem{footnote1} For general $\alpha$, the second line
of Eq.~(\ref{equ:scal_func}) may also read $[\ln(x)]^{s(\alpha)},
s(\alpha)>0$.
\end{thebibliography}
\end{document}